\documentclass[manuscript]{aastex}

\shorttitle{Evolutionary sequences for hydrogen-deficient white dwarfs} 
\shortauthors{Camisassa et al.} 
 
\begin{document} 
 
\title{Updated evolutionary sequences for hydrogen-deficient white dwarfs} 
 
\author{Mar\'ia E. Camisassa, 
        Leandro G. Althaus} 
\affil{Facultad de Ciencias Astron\'omicas y Geof\'isicas, 
       Universidad Nacional de La Plata, 
       Paseo del Bosque s/n, 
       1900 La Plata, 
       Argentina} 
\affil{Instituto de Astrof\'\i sica de La Plata, 
       UNLP-CONICET, 
       Paseo del Bosque s/n, 
       1900 La Plata, 
       Argentina} 
\author{Ren\'e D. Rohrmann} 
\affil{Instituto de Ciencias Astron\'omicas, de la Tierra y del Espacio 
       (CONICET-UNSJ), 
       Av. Espa\~na Sur 1512, J5402DSP, San Juan, 
       Argentina} 
\author{Enrique Garc\'\i a--Berro, Santiago Torres} 
\affil{Departament de F\'\i sica, 
       Universitat Polit\`ecnica de Catalunya, 
       c/Esteve Terrades 5, 
       08860 Castelldefels,  
       Spain} 
\affil{Institute for Space Studies of Catalonia, 
       c/Gran Capit\`a 2-4, 
       Edif. Nexus 201, 
       08034 Barcelona, 
       Spain} 
\and  
\author{Alejandro H. C\'orsico, Felipe C. Wachlin} 
\affil{Facultad de Ciencias Astron\'omicas y Geof\'isicas, 
       Universidad Nacional de La Plata, 
       Paseo del Bosque s/n, 
       1900 La Plata, 
       Argentina} 
\affil{Instituto de Astrof\'\i sica de La Plata, 
       UNLP-CONICET, 
       Paseo del Bosque s/n, 
       1900 La Plata, 
       Argentina}  
 
\begin{abstract} 
We present a set of full  evolutionary sequences for white dwarfs with
hydrogen-deficient atmospheres.  We take into account the evolutionary
history  of the  progenitor  stars, all  the  relevant energy  sources
involved in the  cooling, element diffusion in the  very outer layers,
and outer  boundary conditions provided  by new and  detailed non-gray
white dwarf model atmospheres for pure helium composition. These model
atmospheres are  based on  the most  up-to-date physical  inputs.  Our
calculations extend down to very  low effective temperatures, of $\sim
2\,500$~K, provide  a homogeneous  set of evolutionary  cooling tracks
that  are  appropriate   for  mass  and  age   determinations  of  old
hydrogen-deficient  white dwarfs,  and represent  a clear  improvement
 over   previous  efforts,   which   were   computed  using   gray
atmospheres.
\end{abstract} 
 
 
\section{Introduction} 
\label{introduction} 
 
White dwarfs  are the  most  common fossil  stars within  the stellar
graveyard.  In fact,  it is well known  that more than 90  per cent of
all main  sequence stars will finish  their lives as white  dwarfs. On
their way to become fossil remnants,  half or perhaps even more of the
original  mass  of  main  sequence  stars  is  recycled  back  to  the
interstellar medium,  thus contributing  to its enrichment  in metals.
Supported by electron  degeneracy, white dwarfs continue  to cool down
for  very long  periods of  time, allowing  us to  look back  at early
times. Moreover, their structural  and evolutionary properties are now
reasonably  well  understood ---  see  for  instance, the  reviews  of
\cite{2008PASP..120.1043F},       \cite{2008ARA&A..46..157W}       and
\cite{2010A&ARv..18..471A}   ---   at   least   for   moderately   low
luminosities. Consequently,  the Galactic  population of  white dwarfs
carries  essential   information  about  several   fundamental  issues
\citep{2016NewAR..72....1G}  and is  of crucial  importance to  study,
among other  interesting topics, the  evolution of stars off  the main
sequence, the  structure and evolution  of our  own Galaxy and  of its
components    ---    including    the    thin    and    thick    disks
\citep{1987ApJ...315L..77W, 1988Natur.333..642G,  1999MNRAS.302..173G,
2002MNRAS.336..971T},          the          Galactic          spheroid
\citep{1998ApJ...503..239I,  2004A&A...418...53G}  and the  system  of
open     and      globular     clusters,     see      for     instance
\cite{2010Natur.465..194G},                \cite{2011ApJ...730...35J},
\cite{2013A&A...549A.102B},       \cite{2013Natur.500...51H}       and
\cite{2015A&A...581A..90T}  for  some  examples ---  the  behavior  of
matter     at     high      densities     and     low     temperatures
\citep{1988A&A...193..141G,   1991A&A...241L..29I},   and   also   the
evolution of  planetary systems across  the several phases  of stellar
evolution  \citep{2016NewAR..71....9F}.  Not  only that,  white dwarfs
can   be    also   used   as   astroparticle    physics   laboratories
\citep{1992ApJ...392L..23I, 2008ApJ...675.1512B,   2008ApJ...682L.109I, 
2010A&A...512A..86I, 2012MNRAS.424.2792C}, and to test  a hypothetical secular variation of
the       fundamental      constants       \citep{1995MNRAS.277..801G,
2011JCAP...05..021G, 2013PhRvL.111a0801B}.
 
To undertake these tasks two conditions must be fulfilled. First, from
the observational point  of view we need  observational data to
which the  theoretical models  can be compared.   Surveys such  as the
Sloan  Digital  Sky  Survey  \citep{2000AJ....120.1579Y}  or  the  ESO
Supernovae Type  Ia Progenitor Survey  \citep{2001AN....322..411N}, to
put  only   two  well-known  examples,   have  provided  us   with  an
unprecedented wealth  of information,  allowing us to  make meaningful
comparisons  of  the  theoretical  cooling tracks  with  the  observed
degenerate sequences.  Furthermore, it is foreseen that the results of
future massive surveys,  like the successive data releases  of Gaia or
the  data   obtained  using   the  Large  Synoptic   Survey  Telescope
\citep{2008SerAJ.176....1I}, will revolutionize the research field ---
see, for instance, \cite{2005MNRAS.360.1381T}, and references therein.
However,  being this  important, reliable  cooling sequences  aimed at
reproducing the  characteristics of the observed  populations of white
dwarfs are also needed. In particular, to use white dwarfs as reliable
clocks  to date  stellar populations  precise evolutionary  models are
required.   This, in  turn, means  that all  the relevant  sources and
sinks of energy  must be carefully evaluated, and that  a detailed and
realistic treatment  of the energy  transport in their  atmospheres is
required.   Also, ideally,  a  full end-to-end  treatment  of all  the
stellar evolutionary  phases from  the zero-age main  sequence (ZAMS),
through  the  red  giant  and  the thermally  pulsing  phases  to  the
planetary nebula and cooling stages would be needed.  Only in this way
state-of-the-art  and   realistic  initial  models  for   the  cooling
sequences can be obtained.
 
During the  last decades,  several new  evolutionary models  have been
computed   for    white   dwarfs   with    hydrogen-rich   atmospheres
\citep{1992ApJ...386..539W,  1997ApJ...486..413S, 1998Natur.394..860H,
  1999ApJ...520..680H,    2000ApJ...544.1036S,    2008PASP..120.1043F,
  2010ApJ...716.1241S,    2015A&A...576A...9A,   2016ApJ...823..158C}.
Nevertheless,   about  20   per  cent   of  all   white  dwarfs   have
helium-dominated  atmospheres (the  so-called  class  of non-DA  white
dwarfs). Most  of these white dwarfs  are thought to be  the result of
late  thermal  pulses  experienced  by  post-asymptotic  giant  branch
progenitors \citep{2005A&A...435..631A}.   In contrast with  the large
efforts  devoted   to  model   the  cooling   of  white   dwarfs  with
hydrogen-rich atmospheres  very few works  have been devoted  to study
the  cooling of  white dwarfs  with helium-dominated  atmospheres. The
most  relevant   studies  are  those   of  \cite{1995LNP...443...41W},
\cite{1995PASP..107.1047B},       \cite{1999MNRAS.303...30B}       and
\cite{2009ApJ...704.1605A}.  However, most  of these calculations rely
on  weak grounds.   The reason  for this  is that  the calculation  of
realistic cooling  times for hydrogen-deficient white  dwarfs strongly
depends  on  the  treatment  of the  energy  transport  through  their
atmospheres,  particularly at  low effective  temperatures. The  large
densities that characterize helium white dwarf atmospheres, about $1\,
{\rm g/cm}^3$ in the coolest models, means that non-ideal effects have
to be considered in the radiative  calculations and in the equation of
state.   Thus,  the  modeling  of such  atmospheres  becomes  a  tough
endeavor, and  clearly is a  more complex task than  computing cooling
sequences for white dwarf with hydrogen-rich atmospheres.
 
This paper  is precisely  aimed at filling  this gap.   In particular,
here we  present a suite  of cooling  sequences for white  dwarfs with
pure  helium atmospheres  that  supersede those  sets of  calculations
previously mentioned.  Our initial white dwarf models
are the result of progenitor stars evolved
self-consistently from  the ZAMS through all  the stellar evolutionary
phases, including the thermally pulsing phase and the born again stage. 
The computation of progenitor evolution provides us
with realistic chemical profiles at the beginning of the white dwarf phase.
Our white dwarf sequences incorporate
non-gray detailed atmospheres, and encompass a broad spectrum of masses. 
 
Our paper  is organized  as follows.  In Sect.~\ref{code},  we briefly
describe  our  numerical  tools  and   the  main  ingredients  of  the
evolutionary sequences, the model  atmospheres and the initial models.
In  Sect.~\ref{Res}  we  present  the   results  of  our  white  dwarf
evolutionary sequences, in  particular the cooling times,  the role of
carbon  enrichment in  the envelope,  and the  predicted observational
appearance of  our sequences.  Finally, in  Sect. \ref{conclusions} we
summarize  the  main  findings  of   the  paper  and  we  present  our
conclusions.
 
\section{Numerical setup and input physics} 
\label{code}

\subsection{Stellar code} 
 
The  evolutionary   sequences  presented  in  this   study  have  been
calculated using  the {\tt LPCODE}  stellar evolutionary code  --- see
\cite{2003A&A...404..593A},                \cite{2005A&A...435..631A},
\cite{2012A&A...537A..33A},       \cite{2015A&A...576A...9A},      and
\cite{2016A&A...588A..25M}  for  relevant  information about  the  the
details of the code. {\tt LPCODE} is a well-tested and calibrated code
that has  been widely  used to  study the  formation and  evolution of
white    dwarf     stars    ---     see    \cite{2008A&A...491..253M},
\cite{2010Natur.465..194G},                \cite{2010ApJ...717..897A},
\cite{2010ApJ...717..183R},                \cite{2011ApJ...743L..33M},
\cite{2011A&A...533A.139W},                \cite{2012MNRAS.424.2792C},
\cite{2013A&A...557A..19A},  and references  therein.  More  recently,
the code has been  used to generate a new grid  of models for post-AGB
stars  \citep{2016A&A...588A..25M}.   {\tt  LPCODE}  has  been  tested
against other evolutionary  codes during the main  sequence, red giant
branch,   and    white   dwarf    regime   \citep{2013A&A...555A..96S,
2016A&A...588A..25M} with satisfactory results.
 
In what  follows we describe the  main input physics of  the code that
are   relevant    for   the   computation   of    the   evolution   of
hydrogen-deficient  white dwarfs.   Convection is  treated within  the
standard   mixing   length   formulation,   as  given   by   the   ML2
parameterization \citep{1990ApJS...72..335T}.  The  nuclear network for
helium    burning    is    identical     to    that    described    in
\cite{2005A&A...435..631A}.   In our  sequences, some  residual helium
burning  occurs  at  the  very beginning  of  the  cooling  sequences.
Radiative   opacities    are   computed   using   the    OPAL   tables
\citep{1996ApJ...464..943I}. For the low-temperature regime, molecular
opacities with varying carbon to oxygen ratios are used.  To this end,
we   have  adopted   the   low  temperature   opacities  computed   by
\cite{2005ApJ...623..585F}          and          presented          by
\cite{2009A&A...508.1343W}.   In  our  code, molecular  opacities  are
computed by adopting the opacity tables with the correct abundances of
the  unenhanced  metals  (e.g.   Fe)   and  carbon  to  oxygen  ratio.
Interpolation  is   carried  out   by  means  of   separate  quadratic
interpolations in  $R=\rho/{T_6}^3$ and  $T$, but linearly  in $N_{\rm
C}/N_{\rm  O}$.   Conductive  opacities   are  computed  adopting  the
treatment of \cite{2007ApJ...661.1094C}.  We  consider the equation of
state of \cite{1979A&A....72..134M} for  the low-density regime, while
for  the high-density  regime,  we  employ the  equation  of state  of
\cite{1994ApJ...434..641S}  which  accounts   for  all  the  important
contributions for both the solid  and liquid phases. Neutrino emission
rates  for pair,  photo,  and bremsstrahlung  processes  are those  of
\cite{1996ApJS..102..411I}, while  for plasma processes we  follow the
treatment presented in \cite{1994ApJ...425..222H}.
 
As the white  dwarf cools down we take into  account abundance changes
resulting from  convective mixing  and from  element diffusion  due to
gravitational  settling, chemical  and  thermal  diffusion of  $^4$He,
$^{12}$C,     $^{13}$C,    $^{14}$N     and    $^{16}$O     ---    see
\cite{2003A&A...404..593A} for details.  During the white dwarf regime
and  for  effective  temperatures  smaller  than  $50\,000$  K,  outer
boundary conditions  are derived  from non-gray model  atmospheres for
pure  helium  composition  ---  see  Sect.~\ref{atmospheres}.   Energy
sources resulting from crystallization, as  the release of latent heat
and  of gravitational  energy associated  to carbon  and oxygen  phase
separation upon  crystallization are  also taken into  account.  These
energy sources are considered self-consistently and locally coupled to
the full  set of  equations of stellar  evolution. In  particular, our
treatment of crystallization  includes the use of  the most up-to-date
phase  diagram  \citep{2010PhRvL.104w1101H}   of  dense  carbon-oxygen
mixtures  appropriate for  white dwarf  interiors, which  is based  on
direct      molecular      dynamics      simulations      ---      see
\cite{2012A&A...537A..33A} for the implementation of this treatment in
{\tt LPCODE}.

\subsection{Stellar atmospheres}  
\label{atmospheres} 
 
\begin{figure} 
\centering 
\includegraphics[clip,width=0.99\columnwidth]{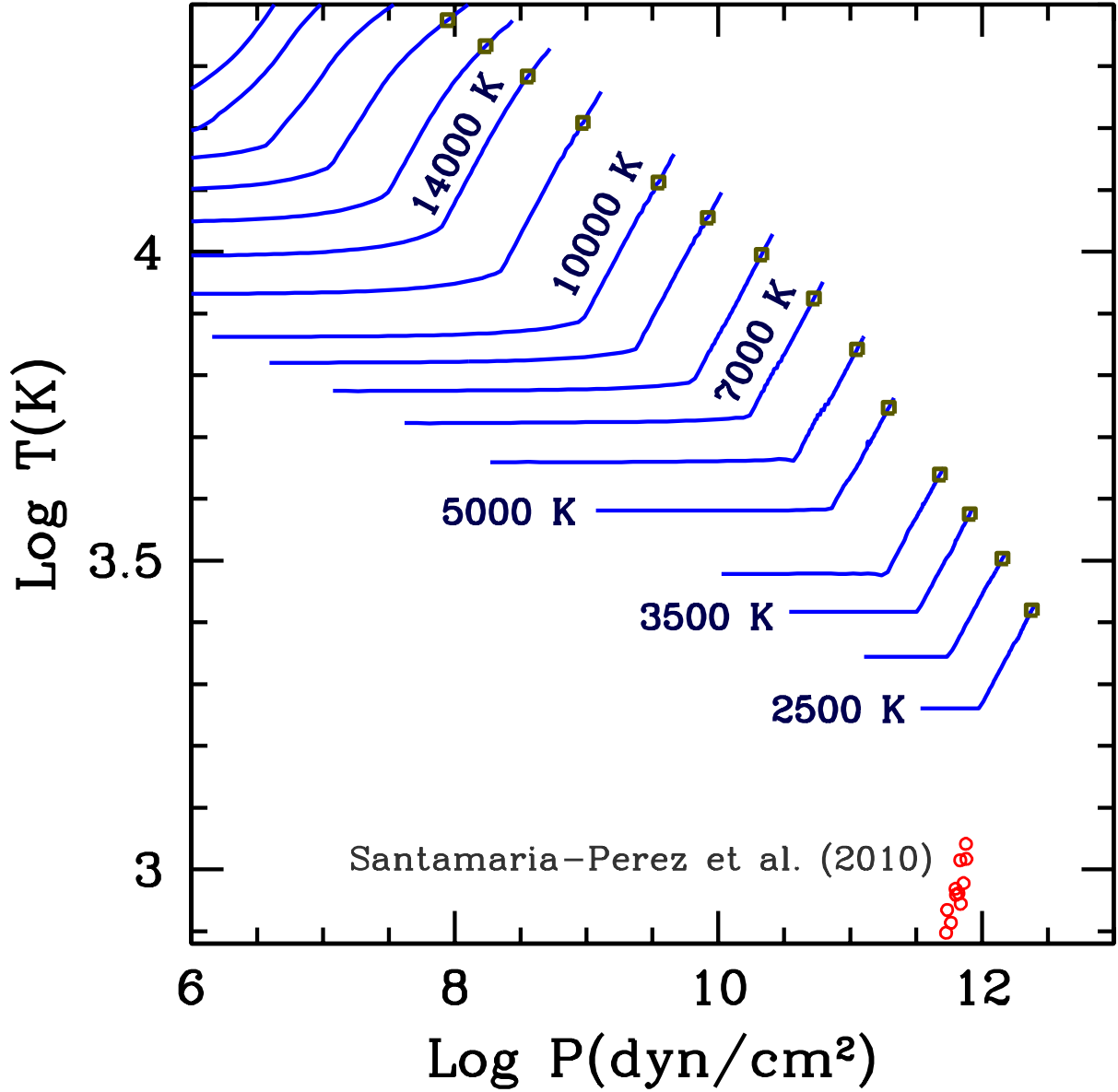} 
\caption{Temperature      and     pressure      stratifications     at
  $10^{-6}\le\tau_{\rm Ross}\le100$ for  helium atmospheres with $\log
  g=8$ and various  effective temperatures (some of  them indicated on
  the   plot).   Squares   display   the   optical  depth   $\tau_{\rm
  Ross}=25.1189$  where  the  atmospheres  are  attached  to  interior
  models.   Circles  show  the  melting data  on  helium  reported  by
  \cite{2010PhRvB..81u4101S}.}
\label{Fig1} 
\end{figure} 
 
Usually,  for hydrogen-deficient  white  dwarfs  the surface  boundary
conditions needed to integrate the  equations of stellar evolution are
taken from the  gray atmosphere approximation.  However, in  the late stages
of  white  dwarf  cooling  precise boundary  conditions  are  required
because convective  coupling between  the envelope and  the degenerate
core occurs  at low effective  temperatures.  In contrast  with nearly
all the calculations  available so far, the  outer boundary conditions
employed  in  this work  are  obtained  from detailed  non-gray  model
atmospheres  for $T_{\rm  eff}\le  50\,000$~K.  The  location of  this
boundary  was taken  at a  large optical  depth (specifically,  at the
Rosseland  mean optical  depth  $\tau_{\rm  Ross}=25.1189$) where  the
diffusion approximation,  limiting form of the  transfer equation used
by the stellar code, is valid --- see \cite{2012A&A...546A.119R}.
 
\begin{figure} 
\centering 
\includegraphics[clip,width=0.99\columnwidth]{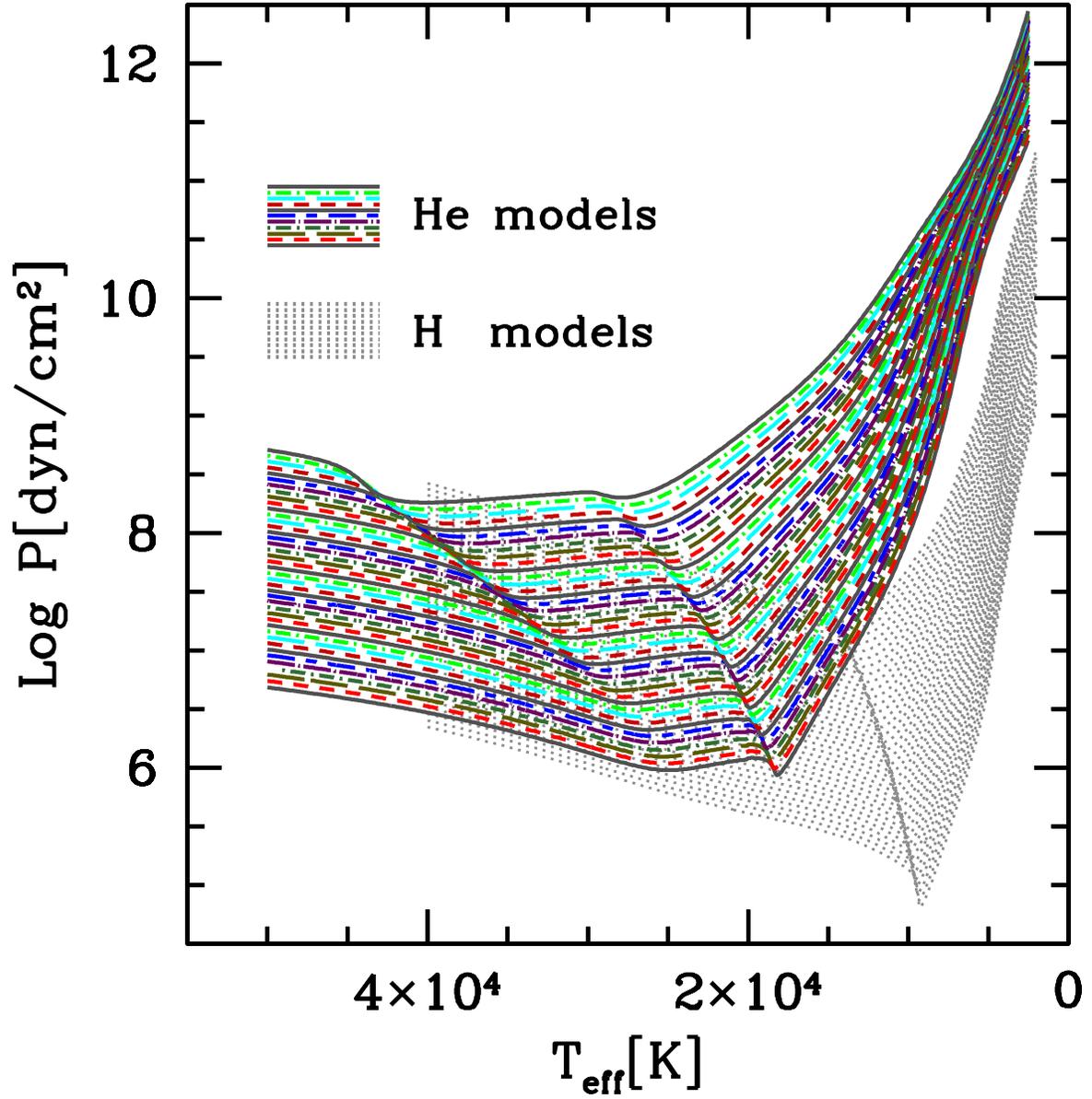} 
\caption{Pressure at  $\tau_{\rm Ross}=25.1189$  as a function  of the
  effective  temperature obtained  for pure  helium (solid  and dashed
  lines) and pure hydrogen (dotted  lines) atmospheres at $5.5\le \log
  g\le 9.5$ (lines from bottom to top).}
\label{Fig2} 
\end{figure} 

Model atmospheres were computed assuming constant gravity, hydrostatic
equilibrium,   local  thermodynamic   equilibrium,  and   energy  flux
conservation.  The numerical  code is a new and improved  version of a
 previously existing code  developed by  \cite{2002MNRAS.335..499R}, appropriately
modified to  take into  account the high  density effects  expected in
cool  helium  atmospheres.    In  the  new  code,   the  equations  of
monochromatic  radiative transfer  and  the  constant flux  condition,
including  convective   and  conductive  energy  fluxes,   are  solved
following  a Rybicki  scheme \citep{1972A&A....19..261G}.   Convective
transport  is treated  within  the ML2  version  of the  mixing-length
theory,   and   the   conductive    opacity   data   is   taken   from
\cite{2007ApJ...661.1094C}.   The  equation  of   state  used  in  the
atmospheric     code      is     the     He.REOS.3      model     from
\cite{2014ApJS..215...21B}, supplemented  with a chemical  model based
on  the  occupation  probability  formalism to  solve  the  ionization
equilibrium for He, He$^+$, He$^{++}$, e$^-$, He$^-$, and He$_2^+$ ---
see  \cite{1988ApJ...331..794H}  and also  \cite{2002MNRAS.335..499R}.
The  radiative   opacity  sources   included  in  the   code  comprise
bound-bound  (He,  He$^+$),  bound-free (He,  He$^+$,  He$^{2+}$)  and
free-free  (He, He$^+$,  He$^{2+}$,  He$^-$)  processes, electron  and
Rayleigh  scattering,   and  collision-induced  absorption   by  atoms
\citep{2014A&A...566L...8K}.   The  density  effects  on  the  opacity
described  in \cite{2002ApJ...569L.111I}  were  omitted  in our  code,
because  unfortunately such  study does  not  cover a  broad range  of
temperatures and  densities as  required in the  present calculations.
However, we have estimated the impact of these high density effects on
the evolutionary timescales, and we  found that the differences in the
cooling times would be less than 2$\%$.
 
Fig.~\ref{Fig1} displays  the temperature-pressure profiles  of helium
atmospheres  for  models  calculated   at  $\log  g=8$  and  different
effective   temperatures.    Temperature  profiles   are   essentially
isothermal in  the outer  and transparent layers  ($\tau_{\rm Ross}\la
10^{-4}$  for  the  coolest  models), while  convection  becomes  very
efficient and the dominant mechanism of energy transport at the bottom
layers ($\tau_{\rm  Ross}\ga 1$) of cool  atmospheres ($T_{\rm eff}\la
20\,000$K). Fig.~\ref{Fig1} also shows  the melting points reported by
\cite{2010PhRvB..81u4101S}, which  correspond to  the liquid  to solid
phase   transition  of   pure   helium.   These   results  show   that
hydrogen-deficient  white  dwarfs   evolve  towards  complex  physical
conditions in their atmospheres as they cool down.
  
\begin{figure} 
\centering 
\includegraphics[clip,width=0.99\columnwidth]{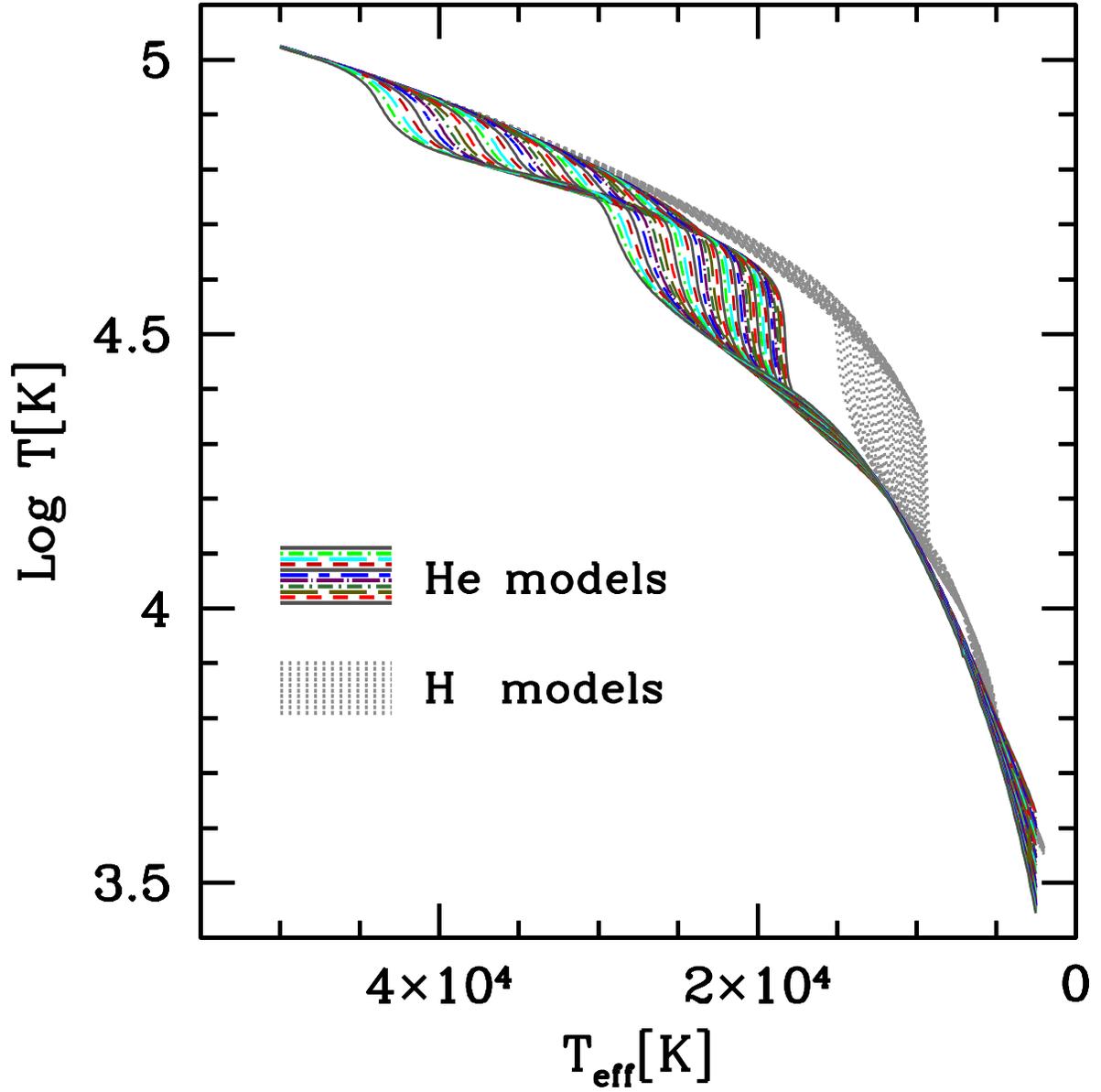} 
\caption{Temperature at $\tau_{\rm Ross}=25.1189$ as a function of the
  effective  temperature obtained  for pure  helium (solid  and dashed
  lines) and pure hydrogen (dotted  lines) atmospheres at $5.5\le \log
  g\le 9.5$ (lines from top to bottom).}
\label{Fig3} 
\end{figure} 

The grid  of atmosphere  models calculated  to provide  outer boundary
conditions  to  the  evolutionary   calculations  covers  a  range  of
effective temperatures from  50\,000~K to 2\,500~K in  steps of 100~K,
and  surface gravities  within  $5.5\le  \log g\le  9.5$  in steps  of
$0.1$~dex.  Pressures and  temperatures (at $\tau_{\rm Ross}=25.1189$)
required to  integrate the  stellar structure  equations are  shown in
Figs.~\ref{Fig2} and \ref{Fig3}, respectively.  Solid and dashed lines
correspond to  our pure-helium  model atmospheres, while  dotted lines
represent data from  pure hydrogen models \citep{2012A&A...546A.119R}.
The pressures  in helium and  hydrogen atmospheres differ  markedly at
low effective temperatures, as a consequence of the differences in the
opacity. The  ripples seen  around $T_{\rm eff}\approx  40\,000$~K and
20\,000~K   in  helium   models  (10\,000~K   in  hydrogen   ones)  in
Figs.~\ref{Fig2} and  \ref{Fig3}, are  a consequence of  He$^{2+}$ and
He$^{+}$ (H$^{+}$) recombinations.
 
\subsection{Initial models}
\label{initial}
 
The  initial models  for our  evolving hydrogen-deficient  white dwarf
sequences  were derived  from  the full  evolutionary calculations  of
their progenitor stars  for $Z=0.02$ \citep{2006A&A...454..845M}.  The
evolution of  the white dwarf  progenitors was computed from  the ZAMS
through the  thermally-pulsing and mass-loss phases  on the asymptotic
giant  branch (AGB),  and finally  to the  born-again stage  where the
remaining hydrogen is violently burned.  After the born-again episode,
the hydrogen-deficient, quiescent remnants of helium burning evolve at
constant  luminosity  with  a  surface chemical  composition  rich  in
helium,    carbon   and    oxygen,    typical    of   PG~1159    stars
\citep{2006A&A...454..845M}.   As  it  will  be shown  below,  such  a
detailed treatment  of the  evolutionary history of  progenitors stars
with   different  initial   stellar  masses   and  distinct   chemical
compositions  is  a  key  ingredient  in  computing  accurate  cooling
sequences  for  cool  hydrogen-deficient white  dwarfs.   The  initial
masses of our sequences and the  white dwarf masses resulting from the
born again episode are listed in Table~\ref{tabla1}. In particular, in
this table  we list,  from left  to right, the  white dwarf  mass, the
initial mass and the total mass of helium of the envelope (all of them
in  solar   units),  and   the  effective  temperature,   the  surface
luminosity, and the surface gravity  at the point of maximum effective
temperature. Finally,  also listed are the  initial surface abundances
(by  mass) of  helium, carbon  and oxygen.   For the  $1.0 \,  M_{\sun}$
initial  model,  two  different   sequences  were  computed  with  two
different mass-loss rates during the AGB evolution.  Thus, we obtain a
different  number of  thermal pulses  and, in  the end,  two different
remnant masses of  $0.515$ and $0.542\, M_{\sun}$.   Finally, in order
to cover  a wide range of  stellar masses, we generated  an additional
$1.0 \, M_{\sun}$ white dwarf  model by artificially scaling our $0.87
\, M_{\sun}$ white dwarf model.  Hence,  the masses of our white dwarf
sequences range  from $0.51$ to  $1.0\, M_{\sun}$. This  mass interval
covers most of  the observed stellar mass  range of hydrogen-deficient
white dwarfs.   Note that for  these sequences, the expected  range of
helium envelope  masses spans  more than one  order of  magnitude.  In
this  work, emphasis  is  placed on  the late  stages  of white  dwarf
evolution, where  the impact of  detailed model atmosphere  is larger.
The evolutionary  stages covering the  hot pre-white dwarf  phases ---
namely, the PG~1159 regime and the  following ones --- were studied in
detail in  \cite{2009ApJ...704.1605A}.  Consequently, we will  not pay
much attention  to the very  early evolutionary phases.   Instead, the
evolution  of these  model  stars  has been  followed  until very  low
surface luminosities,  $\log(L/L_{\sun})=-5.0$, and will  be discussed
in depth here.
 
\begin{table*} 
\centering 
\caption{Main characteristics of our initial white dwarf models.} 
\begin{tabular}{lcccccccc} 
\tableline      
\tableline 
$M_{\rm WD}$ & 
$M_{\rm ZAMS}$& 
$M_{\rm He}$ & 
$\log T_{\rm eff}$ [K] &  
$\log(L/\rm L_\sun)$  &  
$\log g$ [cm~s$^{-2}$] &  
$X_{\rm He}$ &  
$X_{\rm C}$ &  
$X_{\rm O}$ \\  
\tableline 
0.515 & 1.00 & 0.0219 & 5.0634  &  2.6868  &  6.6713 & 0.7437 & 0.1637 & 0.0279  \\ 
0.542 & 1.00 & 0.0072 & 5.1650  &  3.0546  &  6.7412 & 0.2805 & 0.4064 & 0.2127  \\ 
0.584 & 2.50 & 0.0060 & 5.2398  &  3.2574  &  6.8615 & 0.3947 & 0.3060 & 0.1704  \\ 
0.664 & 3.50 & 0.0036 & 5.3578  &  3.3611  &  7.0813 & 0.4707 & 0.3260 & 0.1234  \\ 
0.741 & 3.75 & 0.0019 & 5.4535  &  3.8069  &  7.2701 & 0.4795 & 0.3361 & 0.1390  \\ 
0.870 & 5.50 & 0.0009 & 5.5829  &  4.0961  &  7.5679 & 0.5433 & 0.3012 & 0.0938  \\ 
\tableline 
\tableline 
\end{tabular} 
\label{tabla1} 
\end{table*} 
 
\begin{figure} 
\centering 
\includegraphics[clip,width=0.79\columnwidth]{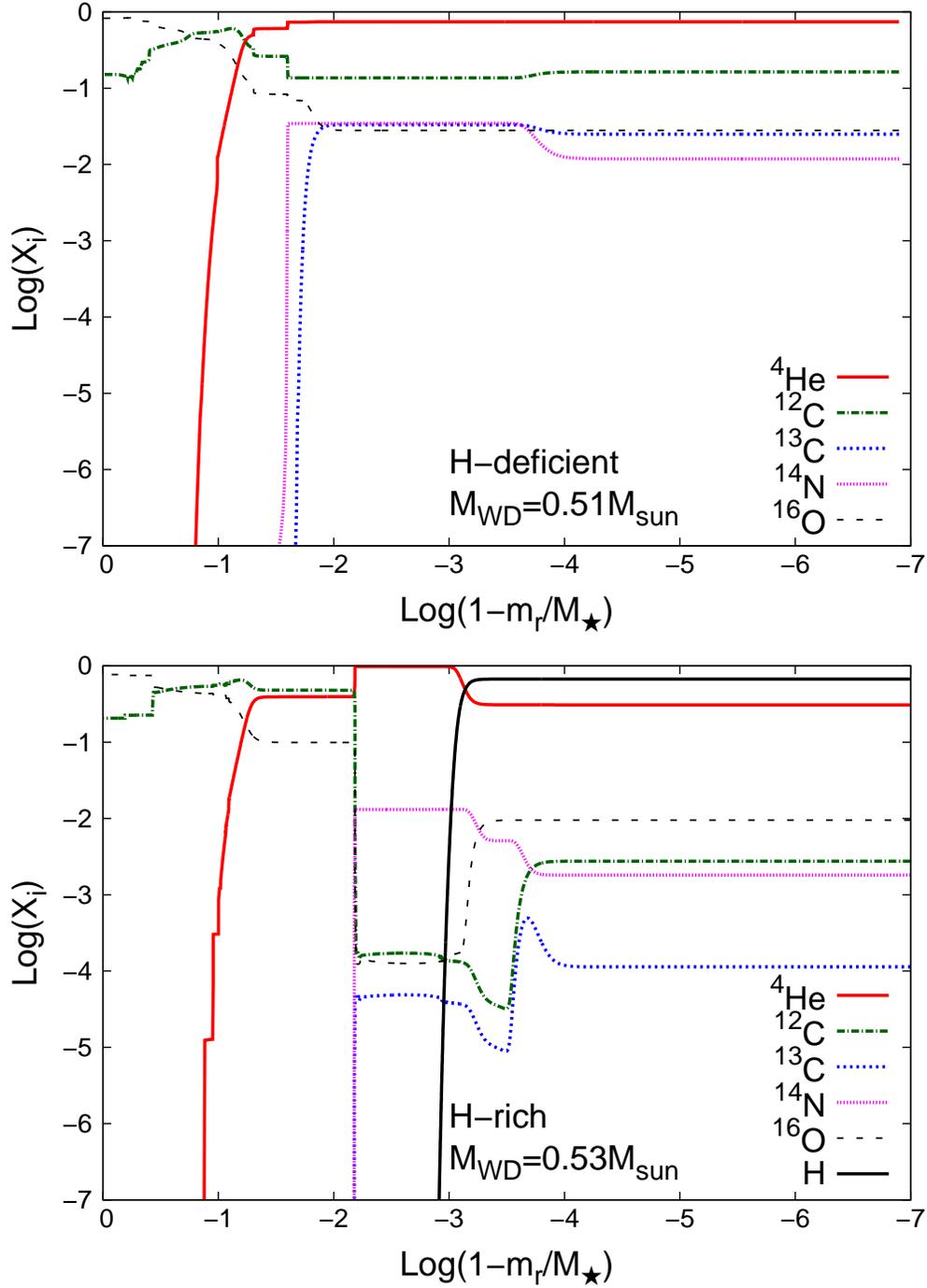} 
\caption{Upper  panel:  initial  chemical  profiles  for  our  $0.51\,
  M_{\sun}$ hydrogen-deficient white dwarf model in terms of the outer
  mass fraction. Bottom panel: initial chemical profiles for a $0.53\,
  M_{\sun}$        hydrogen-rich        white       dwarf        model
  \citep{2016ApJ...823..158C}}
\label{Fig4} 
\end{figure}

\begin{figure} 
\centering \includegraphics[clip,width=0.79\columnwidth]{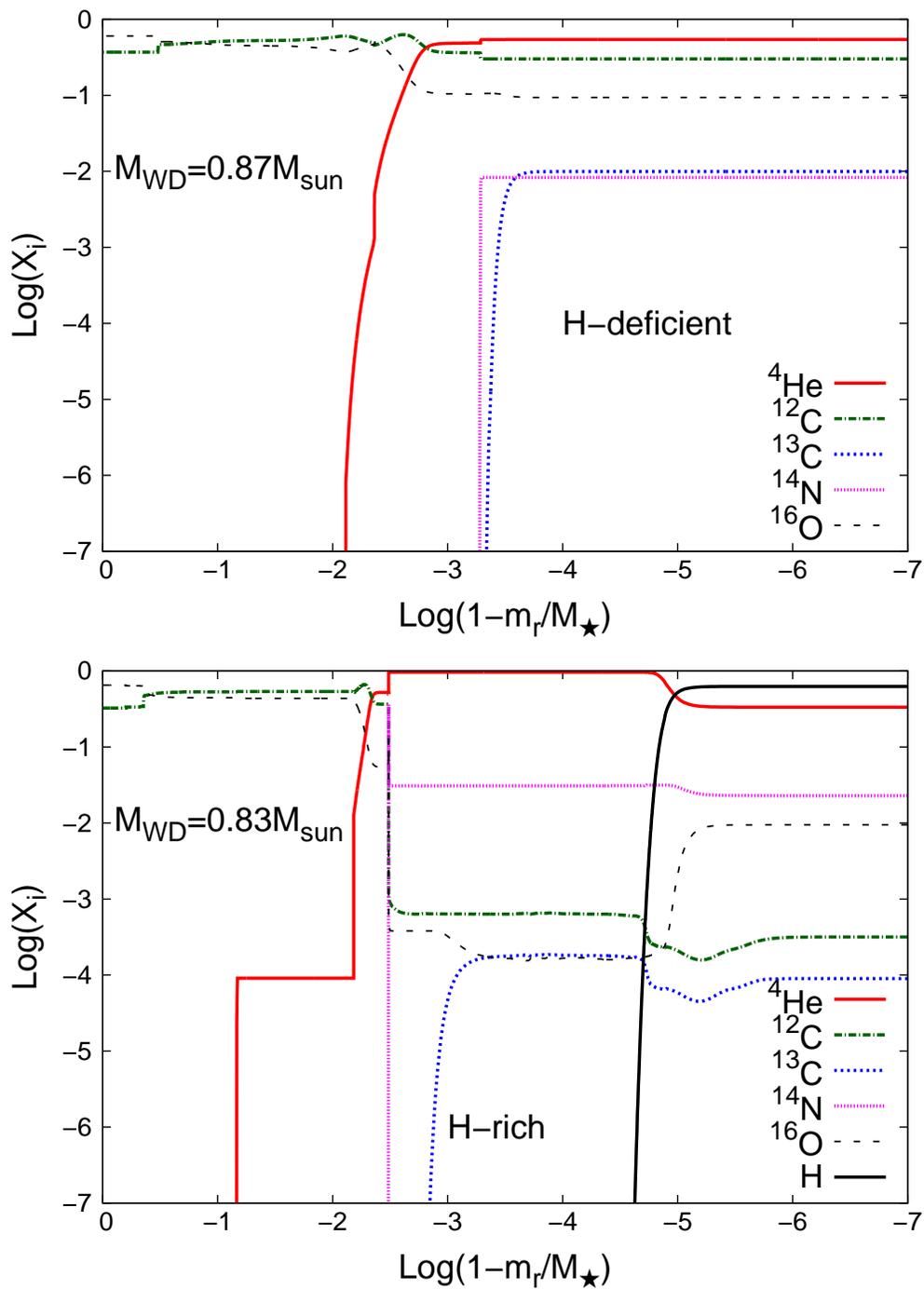}
\caption{Upper  panel:  initial  chemical  profiles  for  our  $0.87\,
  M_{\sun}$ hydrogen-deficient white dwarf model in terms of the outer
  mass fraction. Bottom panel: initial chemical profiles for a $0.83\,
  M_{\sun}$        hydrogen-rich        white       dwarf        model
  \citep{2016ApJ...823..158C}}
\label{Fig5} 
\end{figure}

The initial chemical  abundance distribution (that is, at the beginning of the cooling track)
for some  of our selected
white  dwarf models  is discussed  with the help  of Figs.~\ref{Fig4}  and
\ref{Fig5}, which display  the abundance by mass  of $^4$He, $^{12}$C,
$^{13}$C,  $^{14}$N,  and   $^{16}$O  in  terms  of   the  outer  mass
fraction. In  particular, the upper  panels of these figures  show the
abundances at the beginning of the cooling track of the sequences with
0.515   and  $0.870\,   M_{\sun}$,  respectively.    For  illustrative
purposes,  the inner  chemical  structure  expected for  hydrogen-rich
white dwarf structures of similar stellar  mass is shown in the bottom
panel of each  figure.  Note the dependence of  the chemical abundance
profile on  the stellar mass.   The  signatures left by the  born again
episode in  the outer
layer chemical stratification  is clearly visible. In particular, it can
be seen that hydrogen is almost completely burned in these layers. In  passing we note
that because we are interested  in the evolution of hydrogen-deficient
white dwarfs, we have artificially  removed all the traces of hydrogen
in our initial models.  In particular,  note the high amount of carbon
and  nitrogen  expected in    hydrogen-deficient  white dwarfs,  see
\cite{2006A&A...454..845M} for details. This will be a key issue to be
considered when describing the final  phases of cooling of these white
dwarfs. Note in particular that $^{13}$C reaches abundances as high as
0.05 by mass  throughout the envelope.  It is clear  that, in order to
obtain  realistic  initial  chemical profiles  for  hydrogen-deficient
white dwarfs, the progenitor evolution through the born-again scenario
must  be  taken into  account.  Specifically,  the resulting  chemical
profiles in the  outer layers differ markedly from  those predicted by
progenitors that produce hydrogen-rich white dwarfs.
 
\section{Results} 
\label{Res} 

\begin{figure} 
\centering 
\includegraphics[clip,width=0.95\columnwidth]{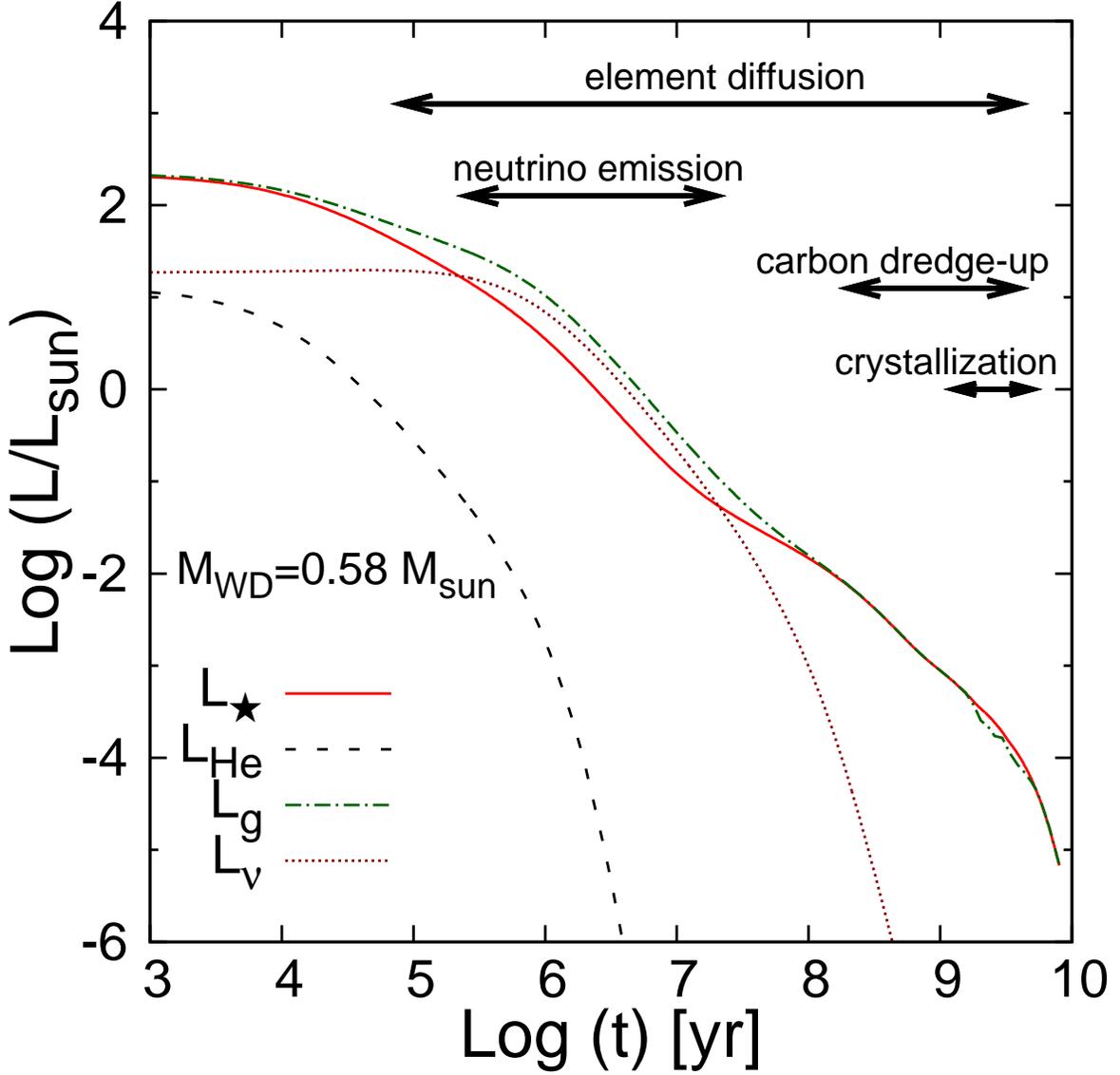} 
\caption{Time dependence of the different luminosity contributions for
  our $0.58 \, M_{\sun}$  hydrogen-deficient white dwarf sequence.  We
  show the surface luminosity,  $L_\star$ (solid line), the luminosity
  due  to helium-burning,  $L_{\rm  He}$ (dashed  line),  the rate  of
  gravothermal (compression plus thermal)  energy release, $L_{\rm g}$
  (dot-dashed  line),   and  the  neutrino  losses,   $L_\nu$  (dotted
  line). Time  is expressed in  years since  the moment when  the star
  reaches  the maximum  effective temperature.   The various  physical
  processes occurring as the white dwarf cools down are also indicated
  in the figure.}
\label{Fig6} 
\end{figure}

A  global view  of  the main  phases  of the  evolution  of a  typical
hydrogen-deficient white  dwarf is  shown in
Fig.~\ref{Fig6}. In this figure the different luminosity contributions
are plotted for  our $0.58\, M_{\sun}$ white  dwarf sequence resulting
from a progenitor  star of $2.5\, M_{\sun}$.  During  the entire white
dwarf evolution,  the release of  gravothermal energy is  the dominant
energy source. Also,  during the very early stages of  the white dwarf
cooling phase, helium  burning contributes  somewhat to  the white dwarf
energy budget. However,  shortly after it becomes  negligible. At
$\log(t)\sim 5.5$ neutrino emission  becomes an important energy sink,
even larger  than the  star luminosity.   At that  time, gravitational
settling has already depleted all the heavy elements from the surface,
leaving an atmosphere   made of almost  pure helium.  However,
gravitational settling is  still acting in the  envelope, and chemical
and  thermal   diffusion  are  still  smoothing   the  inner  chemical
interfaces.  The  resulting chemical stratification will  be discussed
below.  From  $\log(t)\sim 6$  to $7.1$, the  energy lost  by neutrino
emission is of  about the same order of magnitude  as the gravothermal
energy  release.  As  the white  dwarf cools, neutrino  emission   ceases  and,
consequently,  the neutrino  luminosity abruptly  drops.  During  this
phase,  the   outer  convective  zone    grows  inwards   and  at
$\log(t)\sim 8.2$, it  penetrates into layers where  heavy elements as
carbon  and  oxygen  are abundant.   Consequently,  convective  mixing
dredges  up   these  heavy  elements,  and the  surface  composition
changes. In particular, the    outer layers are predominantly enriched
in carbon.   This dredge-up  episode alters markedly  the evolutionary
timescales  of the  white dwarf,  as  it will  be demonstrated  below.
Finally, at $\log(t)\sim  9$ crystallization sets in at  the center of
the  white dwarf.  This  results in  the release  of  latent heat  and
gravitational  energy due  to  carbon-oxygen  phase separation.   Note
that,  as   a  consequence   of  this   energy  release,   during  the
crystallization  phase,  the  surface  luminosity is  larger  than  the
gravothermal  luminosity.   This phase lasts  for  $3.2 \times  10^9$
years.    It  is   also   important  to    note that  during   the
crystallization  phase,  carbon  is  still  being  dredged-up  to  the
surface.   Finally, at  $\log(t)  \sim 9.9$,  the  temperature of  the
crystallized core drops below the Debye temperature, and consequently,
the heat  capacity decreases.  Thus,  the white dwarf enters  into the
so-called ``Debye cooling phase'', where the luminosity drops.
 
\begin{figure} 
\centering 
\includegraphics[clip,width=0.99\columnwidth]{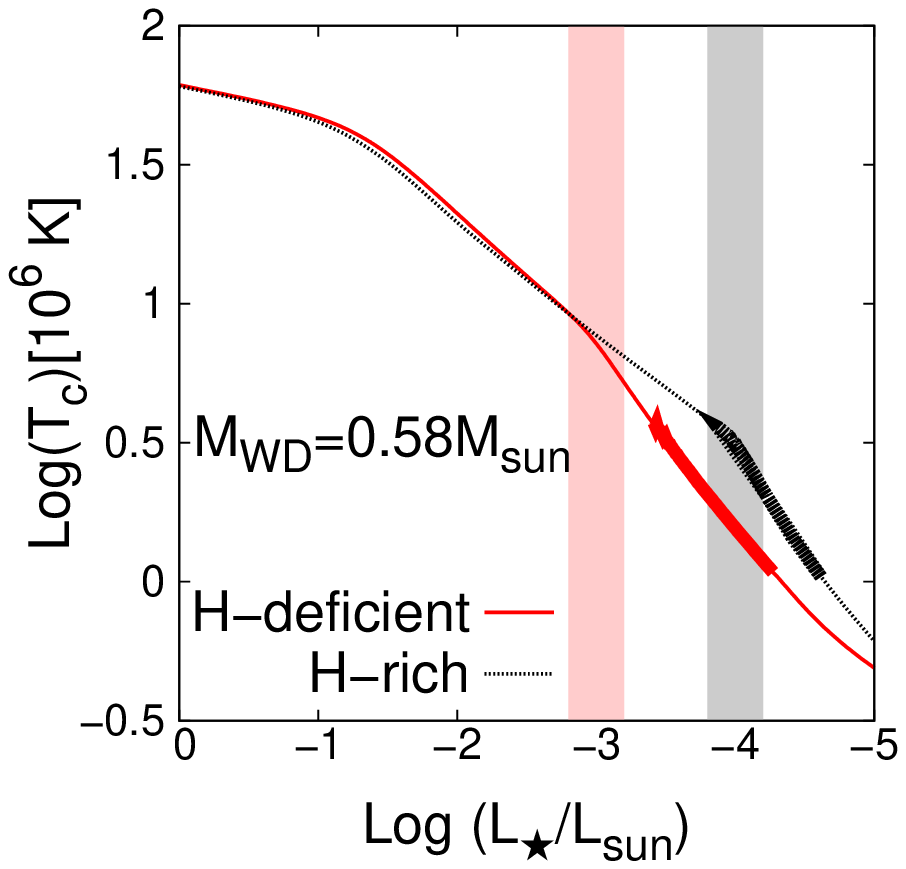}
\caption{Central temperature in  terms of the star  luminosity for our
  $0.58 \,  M_{\sun}$ hydrogen-deficient white dwarf  model (solid red
  line), together with  a $0.58 \, M_{\sun}$  hydrogen-rich model from
  \cite{2016ApJ...823..158C}  (dotted  black  line).   Thick  segments
  indicate the core crystallization  phase.  Shaded areas indicate the
  occurrence of convective coupling.}
\label{Fig7} 
\end{figure}

The behavior  of the cooling sequences  during the late stages  can be
understood by examining Fig.~\ref{Fig7},  which illustrates the run of
the central  temperature in  terms of the  surface luminosity  for our
$0.58  \,  M_{\sun}$ hydrogen-deficient  white  dwarf  model.  In  the
interests  of  comparison,   we  also  include  in   this  figure  the
predictions for  a hydrogen-rich  white dwarf  model of  similar mass.
The thick segments in the  cooling curves indicate the crystallization
phase.  The shaded areas mark the occurrence of convective coupling in
 each sequence, that is, when the outer convective zone reaches
the degenerate core,  and thus the energy transport  through the outer
layers  becomes more  efficient \citep{2001PASP..113..409F}.   For our
hydrogen-deficient  white  dwarf  model, crystallization  sets  in  at
$\log(L/L_\sun)\sim-3.4$,   whereas   in   the   hydrogen-rich   model
crystallization  begins   when  $\log(L/L_\sun)\sim-3.9$.  Crystallization
sets  in at higher luminosities  in hydrogen-deficient
white dwarf  models, and  this  is because  convective coupling  occurs at
higher  luminosities  in  these  white dwarfs.   In  fact,  convective
coupling causes  the change of  slope in  the cooling curves,  and the
core   temperature  strongly   decreases.   This   result,  which   is
independent of  the white dwarf  mass, leads to marked  differences in
the evolution of both types of white dwarfs.
 
\subsection{Chemical evolution} 
\label{CE} 
 
\begin{figure} 
\centering 
\includegraphics[clip,width=0.48\columnwidth]{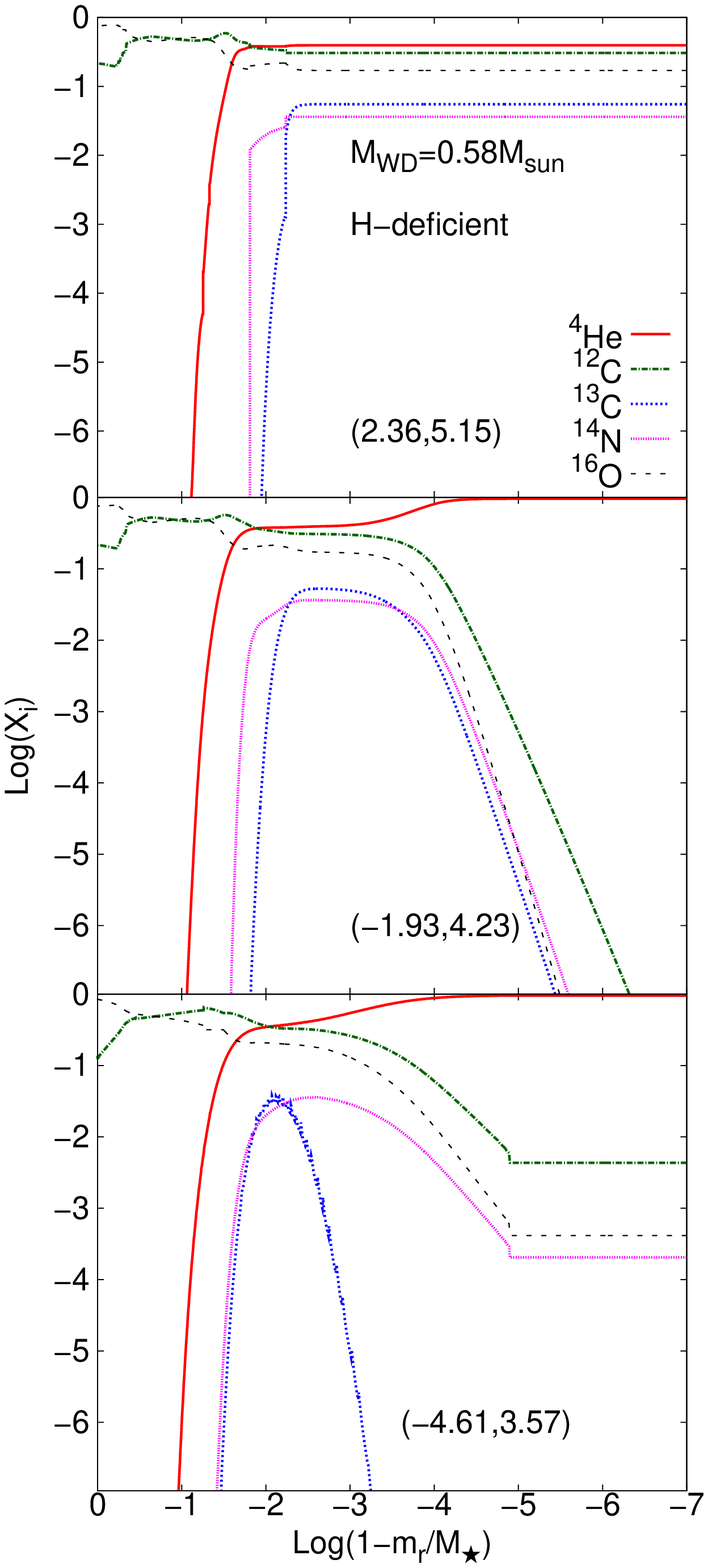} 
\caption{Abundance by mass of $^4$He, $^{12}$C, $^{13}$C, $^{14}$N and
  $^{16}$O   as    a   function   of   the    outer   mass   fraction,
  $\log(1-m_r/M_\star)$, for our $0.58  \, M_{\sun}$ white dwarf model
  at  three  selected  evolutionary   stages.  The  logarithm  of  the
  luminosity  (in solar  units)  and the  logarithm  of the  effective
  temperature are indicated in brackets in each panel.}
\label{Fig8} 
\end{figure} 
 
The chemical profiles at selected evolutionary stages during the white
dwarf  cooling phase  are shown  in Fig.~\ref{Fig8}  for our  $0.58 \,
M_{\sun}$   white dwarf model. Each  panel is  labeled  with the  logarithm of  the
luminosity  (in  solar  units)  and the  logarithm  of  the  effective
temperature of  the evolutionary model.  We show the logarithm  of the
abundances  of $^4$He,  $^{12}$C, $^{13}$C,  $^{14}$N and  $^{16}$O in
terms of the outer mass fraction.   The upper panel corresponds to the
beginning of  the white  dwarf phase.  That is,  we show  the chemical
profiles   at the moment of  maximum  effective  temperature.  As
mentioned,  these  profiles are  the  result  of the  entire  previous
evolution  of the  progenitor  star, particularly  of  the born  again
episode. After  $1.26 \times 10^8$  years, the white dwarf  has cooled
down to $17\,000\, \rm K$ (middle panel), and the chemical composition
displays    the     effects    of    the    action     of    diffusion
processes. Gravitational settling has depleted all heavy elements from
the  outer  layers,  leaving  an envelope  composed  by  pure  $^4$He.
Chemical  and thermal  diffusion  have already  smoothed the  chemical
interfaces. At  this point, the  outer convective zone is  growing and
penetrating into  deeper layers, but  has not yet reached  the regions
where  carbon  is abundant.  Shortly  after,  when $T_{\rm  eff}  \sim
15\,000\, \rm  K$, the outer convective  zone reaches the tail  of the
carbon distribution,  and convection dredges-up carbon  to the surface
layers, increasing the carbon  surface abundance.  Carbon dredge-up by
convection  will  continue  for  $4  \times  10^9$~yr,  until  $T_{\rm
eff}\sim 5\,000\,  \rm K$.  During  this period, thermal  and chemical
diffusion are still acting and further smooth the chemical interfaces,
thus increasing the amount of carbon that is finally dredged-up to the
surface. It  is important  to remark  that convection  also dredges-up
other  chemical species,  but  carbon  is the  one  that is  primarily
dredged-up.  Also,  during this period, crystallization  starts in the
degenerate  core, changing  as  well the  carbon  and oxygen  chemical
profiles  in the  central regions  of  the star.   The final  chemical
abundances,  at $T_{\rm  eff}\sim 3\,700\,  \rm  K$ are  shown in  the
bottom panel of Fig.~\ref{Fig8}.  Note  that at this point, convection
has dredged-up a large amount of carbon to the surface.

\begin{figure} 
\centering 
\includegraphics[clip,width=0.99\columnwidth]{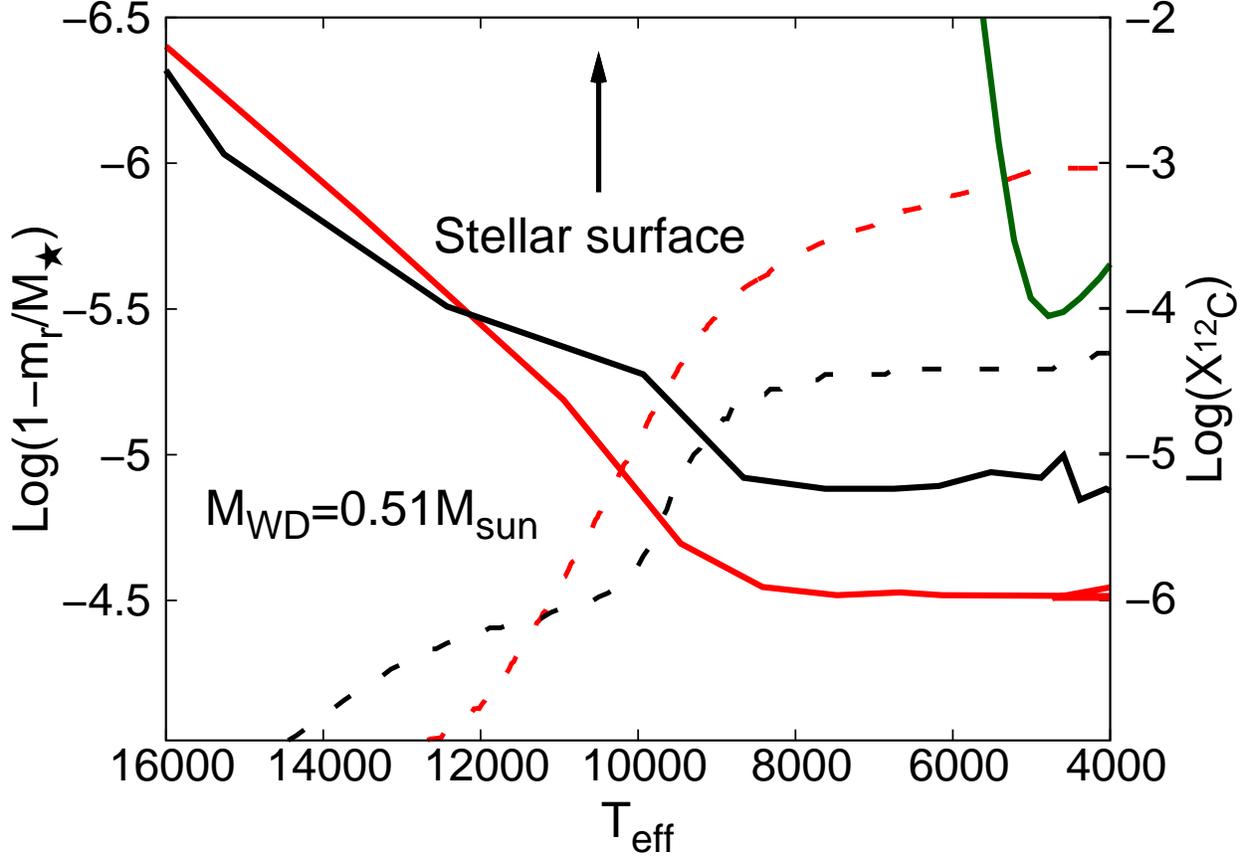} 
\caption{ Effect of convection on the 
carbon abundance in the envelope as a function 
of effective  temperature for  our $0.51  \, M_{\sun}$
  white  dwarf  model.   The  red  and  black  solid  lines  indicate,
  respectively, the base of the outer convective zone resulting from
  considering  a detailed  model  atmosphere and  the Eddington  model
  atmosphere as outer  boundary conditions.  The red  and black dashed
  lines show the surface  carbon abundances resulting from considering
  the  detailed  and  the Eddington  model  atmosphere,  respectively.
  The green  solid line  indicates the  base of  the outer
  convective  zone in  a hydrogen-rich  white dwarf  model of  similar
  mass.  The deeper convection zone of the
detailed model atmosphere results in more efficient dredge-up of carbon and
a higher carbon abundance at the surface.}
\label{Fig9} 
\end{figure} 

For  a  better  understanding  of the  carbon  dredge-up  process,  in
Fig.~\ref{Fig9} we  show using a  red solid  line the location  of the
 base of  the outer convective zone  of our $0.51 \,  M_{\sun}$ white
dwarf sequence in terms of the effective temperature.  For comparison,
we have also  included using a black solid line  the predictions for a
$0.51 \,  M_{\sun}$ hydrogen-deficient white dwarf  sequence resulting
from  considering  the  outer   boundary  conditions  of  an  Eddington
atmosphere model.  As the white dwarf cools, the outer convective zone
penetrates  into  deeper  regions.   At  low  effective  temperatures,
convection  eventually  reaches  deeper  layers in  the  case  that  a
detailed  treatment of  the  atmosphere is  considered, thus  yielding
larger final surface carbon abundances  (dashed red line), as compared
with the  situation in  which the  Eddington atmosphere  is considered
(dashed black  line).  Note also  that for a  hydrogen-deficient white
dwarf the  outer convective  zone penetrates much  deeper in  the star
than in the case in which a hydrogen-rich white dwarf is considered.

\begin{figure} 
\centering 
\includegraphics[clip,width=0.99\columnwidth]{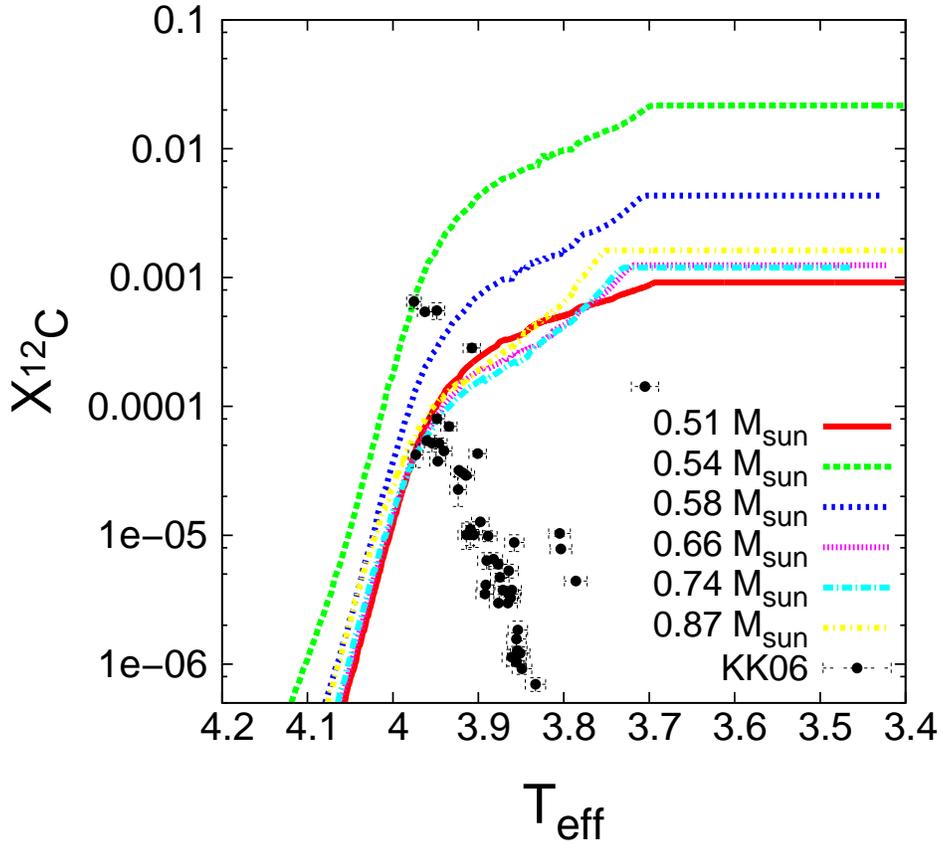} 
\caption{Surface mass fraction  of $^{12}$C in terms  of the effective
  temperature for all our white dwarf sequences.  Filled circles are the DQ white dwarfs
  studied by \cite{2006A&A...454..951K}.}
\label{Fig10} 
\end{figure}

All  our  white  dwarf  models  experience  carbon  dredge-up  due  to
convective mixing. This is  illustrated in Fig.~\ref{Fig10}, where the
surface abundance of carbon of our  sequences is displayed in terms of
the  effective temperature.  All  the white  dwarf cooling  sequences
exhibit  roughly the  same  behavior.  As  a  result of  gravitational
settling, surface  carbon abundance  is negligible  until $\log(T_{\rm
eff})\sim 4.1$.  At  this point, the outer  convective zone penetrates
into  the  carbon-rich layers,  and  consequently  the surface  carbon
abundance increases.  This  process is also favored  by the occurrence
of thermal and chemical diffusion in deeper layers.  When $\log(T_{\rm
eff})\sim 3.7$, the  penetration of the outer  convective zone ceases,
and the surface abundance of  carbon remains essentially constant. The
final surface carbon abundance depends on the stellar mass, as well as
on the initial  chemical abundances. The average  final surface carbon
abundance of  our models  is about  $5 \times  10^{-3}$.  This value is high
enough   to   strongly   modify   the   evolutionary   timescales   of
hydrogen-deficient white dwarfs,  as it will be shown  below.  For the
sake  of comparison,  we also  performed some  additional calculations
considering the  Eddington atmospheres  as outer  boundary conditions,
and  we found that in this  case the resulting evolutionary sequences also
experience appreciable carbon  enrichment in the envelope  as a result
of convective  mixing, although  the carbon  enhancements in  the very
outer layers of the star are substantially smaller.
 Our results can explain  the existence of DQ white dwarfs, which
exhibit traces of carbon in their atmospheres below $\sim 10\,000$K. The white
dwarf models presented in this work undergo a spectral type transition from
DB type to DQ type as their atmospheres become carbon enriched. The filled circles in Fig.~\ref{Fig10}
represent the DQ white dwarfs studied by \cite{2006A&A...454..951K}.

It is important  to emphasize that the total amount  of carbon that is
dredged-up  to  the  surface  is strongly  dependent  on  the  initial
chemical profiles, which are a result of the previous evolution of the
progenitor stars.  Therefore, in order  to correctly assess  the final
surface carbon abundance and to  obtain realistic cooling times, it is
important to  calculate the complete progenitor  evolution through the
thermally pulsing  phase, and more  importantly during the  born again
episode, where the  chemical profiles of carbon  and oxygen throughout
the envelope  that will  characterize the newly  born white  dwarf are
established.  Moreover, chemical and  thermal diffusion also influence
the carbon mixing process, so they must be considered.

\subsection{Evolutionary times}
\label{times}

\begin{figure} 
\centering 
\includegraphics[clip,width=0.99\columnwidth]{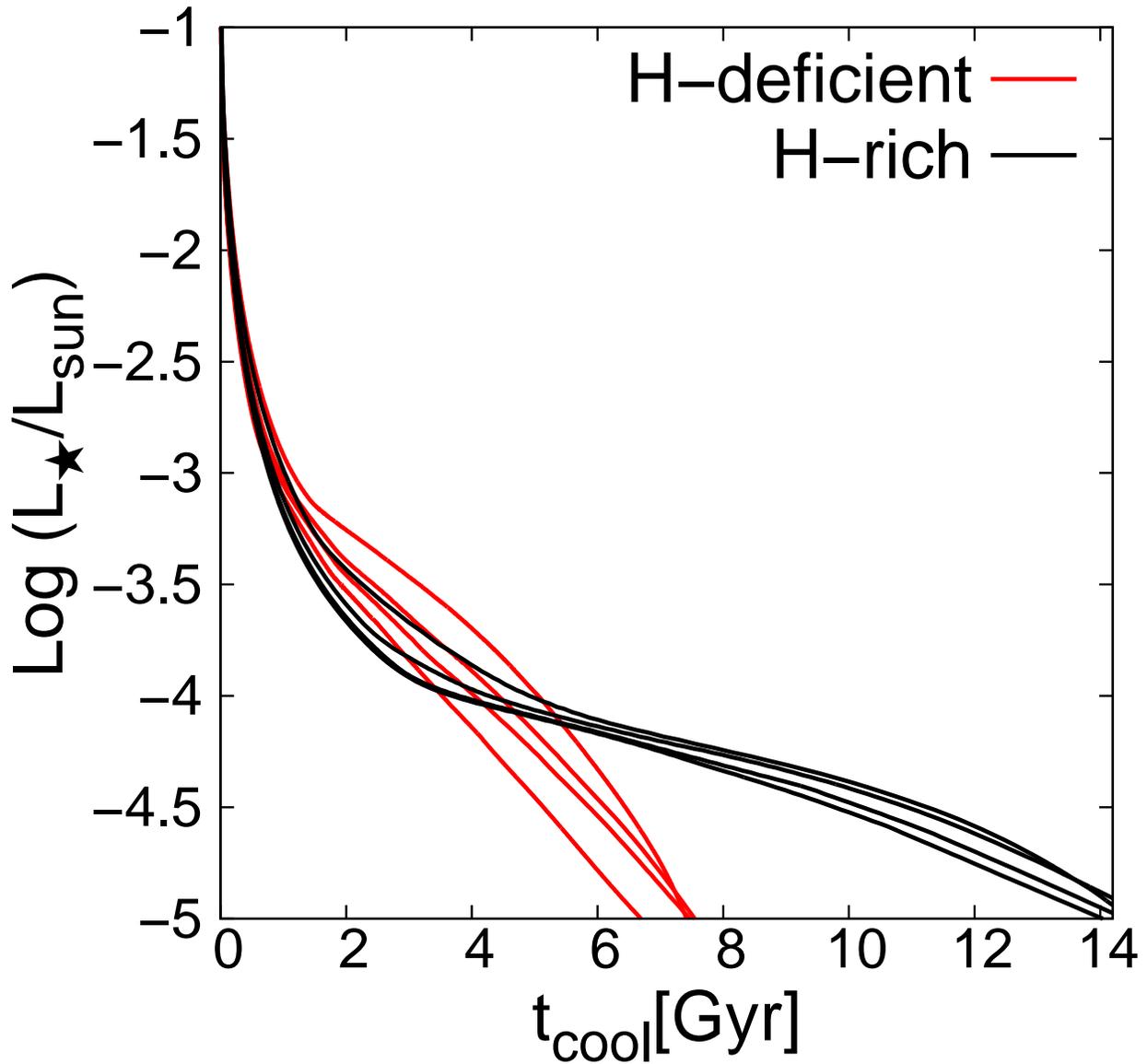} 
\caption{White dwarf evolutionary times.  The red solid lines indicate
  four  hydrogen-deficient white  dwarf sequences  calculated in  this
  work.  At  $t_{\rm cool}=6 \, \rm  Gyr$ and from bottom  to top, the
  different  lines  display the  $0.51$,  $0.58$,  $0.66$ and  $0.87\,
  M_{\sun}$   model  sequences,   respectively.   Black   solid  lines
  correspond to  hydrogen-rich white dwarf tracks  with similar masses
  from \cite{2016ApJ...823..158C}.}
\label{Fig11} 
\end{figure}

The cooling times  for some of  our  sequences are displayed as
red  lines in  Fig.~\ref{Fig11}.  In  addition, we  also plot  in this
figure the  cooling tracks  for hydrogen-rich  white dwarfs of  \cite{2016ApJ...823..158C}, that  have similar  masses to  those
considered in this  work.  The cooling time is defined  as zero at the
beginning of the white dwarf cooling  phase, when the star reaches the
maximum effective temperature.  The cooling times of  all our sequences are
also   listed  in   Table~\ref{tabla2}  for   some  selected   stellar
luminosities.  Some relevant  features of  Fig.~\ref{Fig11} are  worth
commenting.     In   particular,    at   intermediate    luminosities,
hydrogen-deficient  white dwarfs  evolve  markedly  slower than  their
hydrogen-rich  counterparts.  As  explained  previously in  connection
with  Fig.~\ref{Fig7}, this  is because  convective coupling  (and the
associated release  of internal energy) occurs  at higher luminosities
in hydrogen-deficient white dwarfs, with the consequent lengthening of
cooling times at  those luminosities.  In addition  the energy release
resulting  from crystallization  occurs  at higher luminosities  in hydrogen-deficient
white  dwarfs, thus  increasing the  cooling times.   By contrast,  at
low-luminosities, hydrogen-deficient  white dwarfs evolve  faster than
hydrogen-rich  white dwarfs.   This  is because  at  those stages  the
thermal  energy  content of  the  hydrogen-deficient  white dwarfs  is
smaller, and  also because for  these white dwarfs their  outer layers
are  more transparent.  In  fact, note  that hydrogen-deficient  white
dwarfs  evolve  to $\log(L/  L_\sun)\sim-5$  in  less than  8~Gyr,  as
compared with  the time  needed by  hydrogen-rich white  dwarfs to
reach the same luminosity, $\sim 14$~Gyr.

\begin{figure*} 
\centering 
\includegraphics[clip,width=0.7\textwidth]{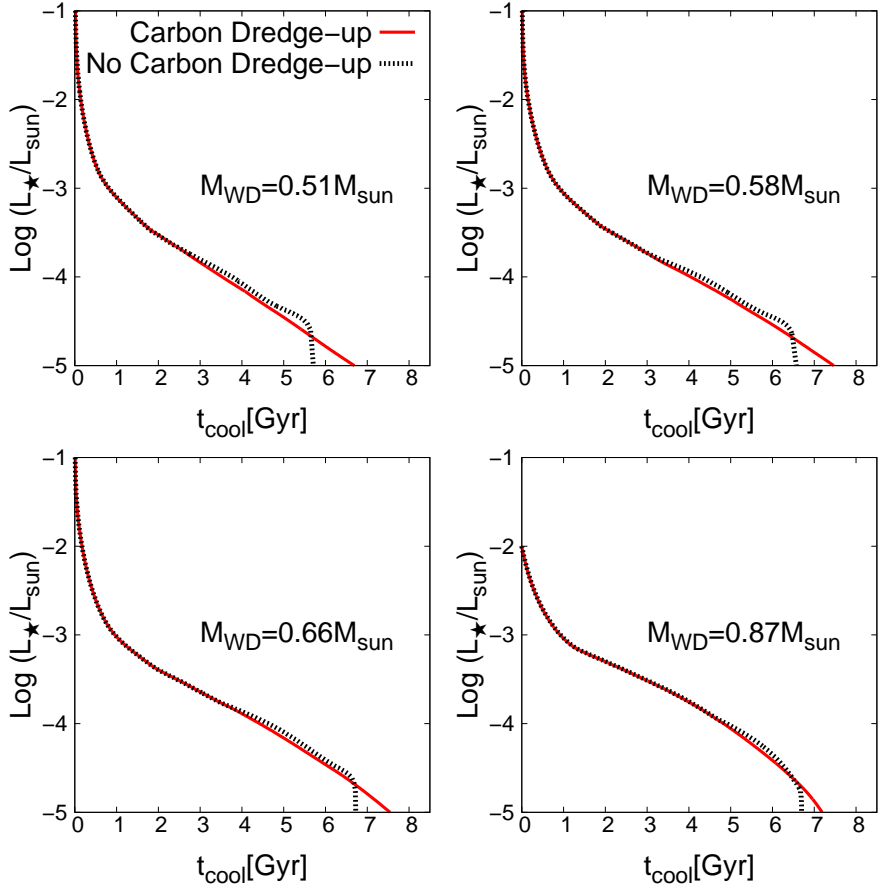} 
\caption {White  dwarf evolutionary times  for the sequences  in which
  carbon is allowed to be dredged-up to the surface by convection (red
  solid lines), and  for the sequences in which  convective mixing was
  suppressed and, therefore, the envelopes are composed of pure helium
  (black dotted lines).  All the sequences have  been calculated using
  our detailed model atmospheres for  pure helium composition as outer
  boundary conditions.}
\label{Fig12} 
\end{figure*}

We remind the reader that  our sequences experience appreciable carbon
dredge-up.    Carbon  enrichment  increases the  opacity  in  the
non-degenerate outer  layers, so energy transport  across these layers
becomes less efficient.  Thus, cooling is delayed and the evolutionary
times increase.   To estimate the  impact of the carbon  enrichment of
the envelope on the white  dwarf evolutionary timescales, we performed
additional evolutionary calculations in which convective mixing in the
envelope was  suppressed, thus  forcing our  white dwarf  envelopes to
remain  composed by pure  helium.  Fig.~\ref{Fig12}  shows the  resulting
cooling times, together with the  cooling times for the sequences with
carbon  contamination in  the  envelopes discussed  in the  preceding
paragraph.   Note that  carbon dredge-up  markedly delays  the cooling
process and increases  the evolutionary times.  These  delays start to
manifest themselves at very low  surface luminosities. In fact, delays
in  cooling times  of  the  order of  1~Gyr  are  reached at  $\log(L/
L_\sun)\sim-5$.
 
\begin{table*} 
\centering 
\caption{Cooling times of our hydrogen-deficient white dwarf models at
  selected luminosities.}
\begin{tabular}{cccccccc} 
\tableline      
\tableline 
$-\log(L/L_\sun)$ & \multicolumn{7}{c}{$t$ (Gyr)}\\ 
\tableline 
 & $0.51 (M_\sun)$ & $0.54\, M_\sun$ & $0.58\, M_\sun$ & $0.66\, M_\sun$ & $0.74\, M_\sun$ & $0.87\, M_\sun$ & $1.00\, M_\sun$ \\ 
\cline {2-8}  
$2.0$ & $0.142$ & $0.136$ & $0.151$ & $0.173$ & $0.193$ & $0.224$ & $0.259$ \\  
$3.0$ & $0.813$ & $0.882$ & $0.909$ & $0.969$ & $1.030$ & $1.148$ & $1.346$ \\ 
$3.5$ & $1.911$ & $2.006$ & $2.161$ & $2.425$ & $2.738$ & $3.165$ & $3.314$ \\ 
$4.0$ & $3.527$ & $3.864$ & $4.056$ & $4.429$ & $4.792$ & $5.055$ & $4.901$ \\ 
$4.5$ & $5.136$ & $5.790$ & $5.890$ & $6.129$ & $6.399$ & $6.438$ & $5.913$ \\ 
$5.0$ & $6.692$ & $7.648$ & $7.496$ & $7.542$ & $7.642$ & $7.419$ & $6.543$ \\ 
\tableline 
\tableline 
\end{tabular} 
\label{tabla2} 
\end{table*} 
 
\begin{figure} 
\centering 
\includegraphics[clip,width=0.86\columnwidth]{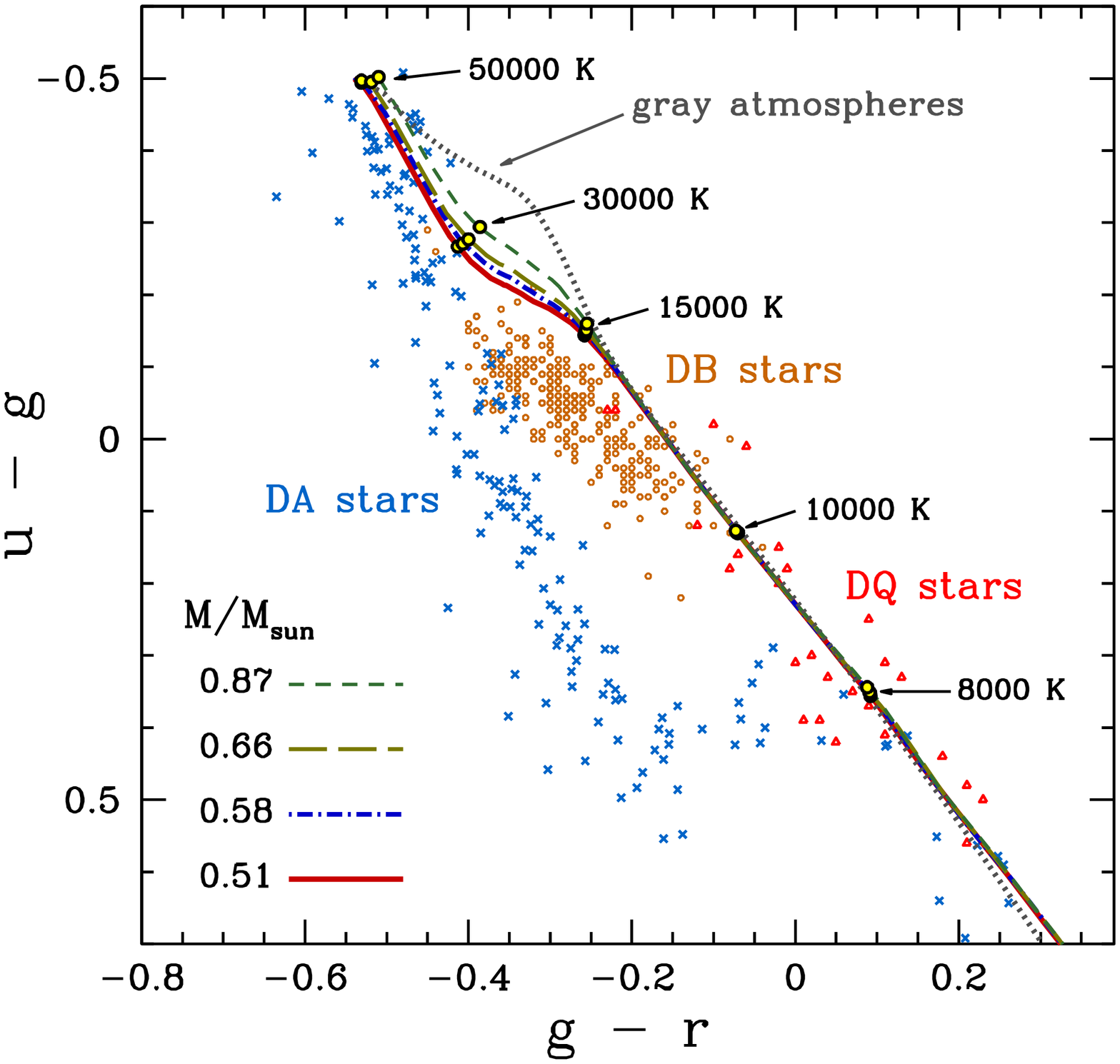} 
\caption{($u-g$)  versus  ($g-r$)  color-color diagram  in  the  SDSS
  photometric  system  for  the  $0.51$, $0.58$,  $0.66$  and  $0.87\,
  M_{\sun}$ cooling  tracks of hydrogen-deficient white  dwarfs.  The
  $0.87\, M_{\sun}$ curve is labeled with effective temperatures. 
  The dotted line displays the same evolutionary tracks calculated  considering the
  gray approximation.
  Open circles
  are  the DB white dwarfs studied  by
  \cite{2015A&A...583A..86K}, open triangles are the DQ white dwarfs studied
  by \cite{2003AJ....126.1023H} and \cite{2004ApJ...607..426K}, 
  and crosses  are the DA  white  dwarfs  studied
  by \cite{2006AJ....132.1221H}.}
\label{Fig13} 
\end{figure}

The observational appearance of our  theoretical cooling tracks can be
obtained from  the radiative transfer calculations  performed with the
atmospheric numerical  code.  Fig.~\ref{Fig13} presents  a color-color
diagram in  the Sloan  Digital Sky  Survey (SDSS)  photometric system,
showing  the cooling  tracks of  hydrogen-deficient white  dwarfs with
masses $0.51$,  $0.58$, $0.66$ and $0.87\,  M_{\sun}$.  Observed white
dwarfs   with   spectroscopically    determined  pure  helium   atmospheres
\citep{2015A&A...583A..86K} --- that  is, of the  DB spectral type  --- are
plotted using open circles.  Observed white dwarfs with atmospheres composed by helium with
traces of carbon \citep{2003AJ....126.1023H, 2004ApJ...607..426K} ---  that is,
of  the DQ spectral  type --- are plotted using open
triangles. Those  white dwarfs with  hydrogen atmospheres
\citep{2006AJ....132.1221H} ---  that is, of  the DA spectral  type ---
are represented using crosses.  There  is a considerable scatter in the
observed  colors of  these stars,  with the  DB and DQ white dwarfs  located in  the
middle of the diagram, and
DA white  dwarfs located at the left of the helium  sequences. 
 Note  that  the theoretical  colors of our
white  dwarf models increase  steadily  along  the
evolution. The theoretical colors show a moderate dependence
on  the   stellar  mass  for  effective   temperatures  above  $T_{\rm
eff}\approx 15\,000$~K,  while all  sequences fall on  a same  line at
lower  effective  temperatures.   It should be emphasized that our white dwarf sequences
change their spectral type from DB to DQ at $T_{\rm
eff}\approx 15\,000$~K as a result of carbon dredge-up due to convective mixing.
The  agreement  with  the
observational data is quite good.   Also, for the sake of comparison, we have calculated
the observational appearance of the same cooling tracks but considering the gray approximation (dotted line).
The effect of solving the radiative transfer for each frequency is obvious in the
regions where the helium opacity has a strong dependence on the wavelength, this is for
$T_{\rm
eff}\gtrsim 15\,000$~K and for $T_{\rm
eff}\lesssim 7\,000$~K.
We finally stress that our model  atmospheres
represent a  considerable improvement  over the  gray atmospheres
that most models employ.

\section{Summary and conclusions} 
\label{conclusions} 
 
We have computed evolutionary cooling sequences for hydrogen-deficient
white dwarfs  resulting from  Solar metallicity progenitors,  aimed at
providing reliable  evolutionary cooling tracks for  these stars.  For
that  purpose, in  our  calculations we  considered detailed  non-gray
model atmospheres  made of pure  helium. These model  atmospheres take
into account  the most advanced prescriptions  of high-density effects
on the radiative processes and in  the equation of state, as described
in   Sect.~\ref{atmospheres}.    Moreover,   for   our   white   dwarf
evolutionary calculations also we took into account the predictions of
full evolutionary calculations of  the corresponding progenitor stars,
as explained in Sect.~\ref{initial}.  These calculations encompass the
full history of the progenitor star, since they start at the ZAMS, and
were evolved through the thermally  pulsing AGB and, most importantly,
through  the born-again  scenario, which  dictates the  final chemical
composition of the outer layers of our model white dwarfs.  Therefore,
the  initial  chemical  profiles  of our  evolving  white  dwarfs  are
self-consistent  and  realistic,  as  they result  from  the  previous
evolution  of  the progenitors  of  white  dwarfs.  Moreover,  these
chemical profiles were evolved
 taking  into account  element  diffusion.   Finally, the  mass
interval of  our cooling sequences  spans from $\sim 0.5\,  M_\sun$ to
$\sim 1.0\, M_\sun$, and thus cover the entire range of masses typical
 of carbon-oxygen white dwarfs.  In short,  the set of white dwarf cooling
tracks presented here  is the first suite  of homogeneous evolutionary
calculations of hydrogen-deficient white dwarfs in the literature that
consider non-gray  atmospheres, include in detail  the density effects
on  the  energy  transport  in the  cool  helium  atmosphere,  element
diffusion,   consider  the   entire   evolutionary   history  of   the
corresponding progenitor stars, and cover  the full range of masses of
interest. Consequently, they 
can be considered as  an important step
forward  in   this  kind  of   calculations\footnote{The  evolutionary
sequences   presented   in   this   work  can   be   downloaded   from
http://evolgroup.fcaglp.unlp.edu.ar.}.
 
Our  calculations  show  that  convective coupling  occurs  at  larger
luminosities in hydrogen-deficient white  dwarfs than in hydrogen-rich
white  dwarfs. As  a  result  of the  associated  release of  internal
energy, at  intermediate-luminosities hydrogen-deficient  white dwarfs
evolve slower  than hydrogen-rich white dwarfs.  At lower luminosities
the reverse  occurs.  In  particular, hydrogen-deficient  white dwarfs
reach  luminosities  as low  as  $\log(L/L_\sun)\sim-5$  in less  than
8~Gyr,  whereas hydrogen-rich  white dwarfs  reach this  luminosity in
about 14~Gyr, see Sect.~\ref{times}.
  
It is  important to  highlight that in  all our  hydrogen-deficient white
dwarf sequences, at $T_{\rm eff}  \sim 15\,000$K, the outer convective
zone reaches the carbon-rich layers,  and the surface carbon abundance
increases as a result of convective mixing.
The  chemical enrichment in carbon of the very  outer
  layers of our model stars has a direct consequence, as the spectral type
  of our model white dwarfs  evolves from  DB  to DQ.  This is consistent with the properties of the observed 
white dwarf population. We calculated the changes
in  the  chemical abundances  due  to  convective mixing  and  element
diffusion,  coupled with  the  thermal evolution  of  our white  dwarf
models. We found  that the carbon enrichment is larger  when the outer
boundary  conditions  are  those   predicted  by  our  detailed  model
atmospheres compared  to the case  in which Eddington  atmospheres are
considered.  We also found that (quite naturally) the cooling times of
hydrogen-deficient  white  dwarfs  depend  on  the  amount  of  carbon
dredged-up  to  the  surface  by convection.   In  particular,  carbon
dredge-up results in  marked delays on the evolutionary  times at very
low surface luminosities.  In this sense, it is worth emphasizing that
detailed  initial chemical  profiles in  the envelope  of these  white
dwarfs, as predicted by evolution of their progenitor stars during the
last thermal pulse, are crucial.  Moreover, our prediction
  that the outer layers are enriched in carbon, implies that the chemical stratification predicted
by evolution of their progenitor stars is in line with the carbon enrichment inferred 
from observations. 
 
Our  results also  show  that detailed  helium  model atmospheres  are
needed to compute reliable  cooling times for hydrogen-deficient white
dwarfs.  Although our cooling  sequences represent a clear improvement
over previous efforts in computing the evolution of such white dwarfs,
a cautionary remark  is in order at this point.   As mentioned, at low
effective   temperatures,    the   pristine   helium    envelopes   of
hydrogen-deficient   white  dwarfs   become   convective,  and   their
atmospheres are contaminated  by the dredged-up material.   This has a
consequence  that   our  models  do  not   consider.   Indeed,  carbon
contamination in the  atmosphere, of course, results  in an additional
source of  opacity, and consequently  the atmospheres of  white dwarfs
made of almost  pure helium become more opaque.  However,  to the best
of  our  knowledge,  calculations  of   the  carbon  opacity  for  the
thermodynamic conditions  prevailing in such dense  atmospheres do not
exist yet.  Consequently, in this first set of calculations we did not
attempt to incorporate  the effects of the enhanced  carbon opacity on
the model atmospheres.  Hence, our  evolutionary cooling times at very
low  effective  temperatures  should   be  taken  with  some  caution.
However,  as   mentioned,  for   larger  effective   temperatures  our
calculations are  clearly superior  to those  existing now.   Thus, we
defer  the  calculation of  improved  cooling  sequences at  very  low
effective temperatures  --- taking into  account the effects  of metal
contamination ---  to forthcoming works.

\acknowledgments 
 We warmly thank M. M. Miller Bertolami for a useful discussion that helped to
improved the final version of the paper.
Part  of this work  was supported by  AGENCIA through
the Programa  de Modernizaci\'on Tecnol\'ogica BID  1728/OC-AR, by the
PIP 112-200801-00940  and PIP 112-200801-01474 grant  from CONICET, by
MINECO grant  AYA2014-59084-P, and  by the  AGAUR.  This  research has
made use of NASA Astrophysics Data System. 

\bibliographystyle{apj}
\bibliography{db} 
 
\end{document}